\documentclass[twocolumn]{aastex62}

\graphicspath{{./}{figures/}}

\usepackage{todonotes}
\usepackage{aas_macros}
\usepackage{graphicx}
\usepackage{dcolumn}
\usepackage{bm}
\usepackage{tikz}
\usepackage{amsmath,amssymb}

\usepackage [autostyle, english = american]{csquotes}
\MakeOuterQuote{"}

\submitjournal{ApJ}

\shorttitle{GR and the trouble with $H_0$}
\shortauthors{Macpherson et al.}

\begin{document}

\title{The trouble with Hubble: Local versus global expansion rates in inhomogeneous cosmological simulations with numerical relativity}

\correspondingauthor{Hayley J. Macpherson}
\email{hayley.macpherson@monash.edu}

\author{Hayley J. Macpherson}
\affil{Monash Centre for Astrophysics and School of Physics and Astronomy, \\
Monash University, VIC 3800, Australia}

\author{Paul D. Lasky}
\affiliation{Monash Centre for Astrophysics and School of Physics and Astronomy, \\
Monash University, VIC 3800, Australia}
\affiliation{OzGrav: The ARC Centre of Excellence for Gravitational-wave Discovery, Clayton, VIC 3800, Australia}

\author{Daniel J. Price}
\affiliation{Monash Centre for Astrophysics and School of Physics and Astronomy, \\
Monash University, VIC 3800, Australia}



\begin{abstract}
In a fully inhomogeneous, anisotropic cosmological simulation performed by solving Einstein's equations with numerical relativity, we find a local measurement of the effective Hubble parameter differs by less than 1\% compared to the global value. This variance is consistent with predictions from Newtonian gravity. We analyse the averaged local expansion rate on scales comparable to Type 1a supernova surveys, and find that local variance cannot resolve the tension between the \citet{riess2018b} and \citet{planck2018a} measurements.

\end{abstract}



\section{Introduction} \label{sec:intro}
Recently, the tension in the locally measured value of the Hubble parameter, $H_0$ \citep{riess2011,riess2016} and that inferred from the cosmic microwave background (CMB) \citep{planck2018a} has reached $3.6\sigma$ \citep{riess2018a,riess2018b}. This tension has both motivated the search for extensions to the standard cosmological model, and for the improvement of our understanding of systematic uncertainties \citep[e.g.][]{efstathiou2014,addison2016,dhawan2018a}. The higher local expansion rate \citep{riess2018a,riess2018b} suggests we may live in a void \citep{cusin2017,sundell2015}, consistent with local $\sim20-40\%$ underdensites that have been found in the supernovae Type 1a (SNe) data \citep{zehavi1998,jha2007,hoscheit2018}. 

In an attempt to address this tension, we perform cosmological simulations of nonlinear structure formation that solve Einstein's equations directly with numerical relativity. In this letter we quantify local fluctuations in the Hubble parameter based purely on physical location in an inhomogeneous, anisotropic universe. Further details of our simulations are given in \citet{macpherson2018a}, including a quantification of backreaction of inhomogeneities on globally averaged quantities.

Local fluctuations in the expansion rate due to inhomogeneities have been analysed using Newtonian and post-Friedmannian N-body cosmological simulations \citep[e.g.][]{shi1998,wojtak2014,odderskov2014,odderskov2016,adamek2017}, second-order perturbation theory \citep{ben-dayan2014}, and exact inhomogeneous models \citep[e.g.][]{marra2013}. These approaches predict local fluctuations in the Hubble parameter of up to a few percent. Inhomogeneities have also been proposed to have an effect on the globally measured expansion rate \citep[e.g.][]{buchert2015,roy2011}, with analytical approaches showing this can contribute to an accelerated expansion \citep[e.g.][]{rasanen2006b,rasanen2008,ostrowski2013}. Under the "silent universe" approximation, a globally, non-flat geometry has been shown to fully alleviate the Hubble tension \citep{bolejko2017b,bolejko2018b}. These works are important steps towards fully quantifying the effects of inhomogeneities on the Hubble expansion, although simplifying assumptions about the inhomogeneities themselves limit the ability to make a strong statement.

Considering a fully inhomogeneous, anisotropic matter distribution in general relativity allows us to analyse the effects of inhomogeneities without simplifying the structure of the Universe. Simulations of large-scale structure formation with numerical relativity have been shown to be a viable way to study inhomogeneities \citep{giblin2016a,bentivegna2016a,macpherson2017a,giblin2017a,east2018}, although fluctuations in the Hubble parameter have not yet been considered. In this work we attempt to quantify the discrepancy between local and global expansion rates using cosmological simulations performed without approximating gravity or geometry.

We present our computational and analysis methods in Section~\ref{sec:method}, and outline our method for calculating the Hubble parameter in Section~\ref{sec:hubble}. We present results in Section~\ref{sec:results} and discuss them in Section~\ref{sec:discuss}.

Redshifts quoted throughout this paper are based purely on the change in conformal time, and are stated as a guide to the reader, rather than corresponding to an observational measurement. We adopt geometric units with $G=c=1$, unless otherwise stated. Greek indices run from 0 to 3, and Latin indices run from 1 to 3, with repeated indices implying summation.

\section{Method} \label{sec:method}
We have simulated the growth of large-scale cosmological structures using numerical relativity. Our initial conditions were drawn from temperature fluctuations in the CMB radiation, using the Code for Anisotropies in the Microwave Background \citep[CAMB;][]{lewis2002}. The initial density perturbation is a gaussian random field drawn from the matter power spectrum of the CMB\footnote{To create a gaussian random field following a particular power spectrum, we use the Python module \texttt{c2raytools}: https://github.com/hjens/c2raytools}, and the corresponding velocity and spacetime perturbations were found using linear perturbation theory. We use the free, open-source Einstein Toolkit along with our thorn \texttt{FLRWSolver} \citep{macpherson2017a} for defining initial perturbations. In a previous paper we benchmarked our computational setup for homogeneous and linearly perturbed cosmological solutions to Einstein's equations, achieving precision within $\sim10^{-6}$ \citep[see][]{macpherson2017a}. For full details of our computational methods, including generation of initial conditions, derivations of the appropriate equations, details of gauge and more we refer the reader to \citet{macpherson2018a}.

\begin{figure*}
	\includegraphics[width=\textwidth]{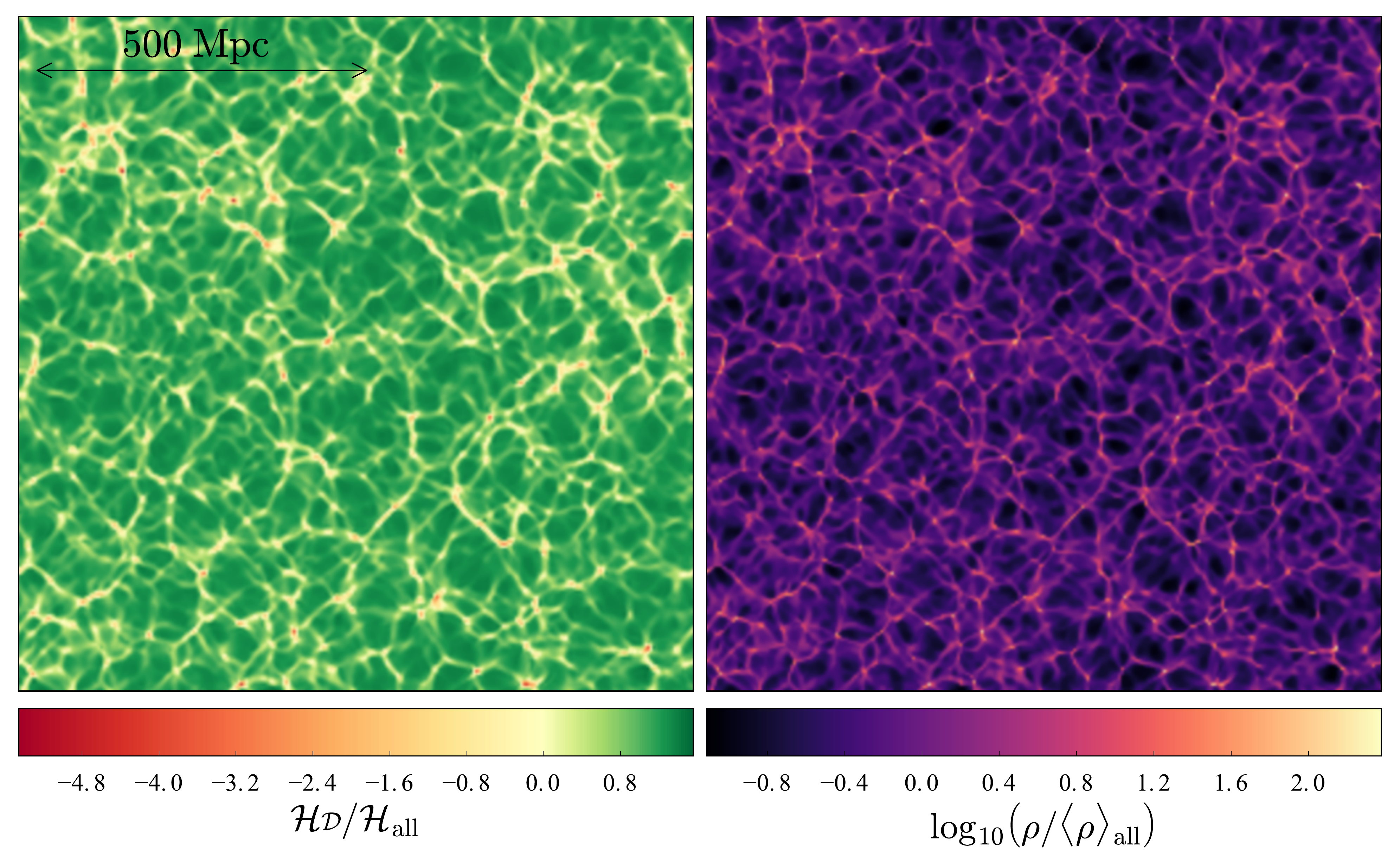}
    \caption{Expansion rate and density of an inhomogeneous, anisotropic universe. Left panel shows the deviation in the Hubble parameter relative to the global mean $\mathcal{H}_\mathrm{all}$. Right panel shows the density distribution relative to the global average, $\langle\rho\rangle_\mathrm{all}$. Both panels show a slice through the midplane of a $256^{3}$ resolution simulation with $L=1$ Gpc.}
    \label{fig:thetarho}
\end{figure*}

We evolve Einstein's equations in full, with no assumed background cosmology, beginning in the longitudinal gauge from $z=1100$, through to $z=0$. Since we have not yet implemented a cosmological constant in the Einstein Toolkit, we assume $\Lambda=0$, and a matter-dominated ($P\ll\rho$) universe. This implies the age of our model universe will differ from the Universe where $\Lambda\neq0$. We simulate a range of resolutions and domain sizes, detailed in \citet{macpherson2018a}. Here we analyse a $256^{3}$ resolution, $L=1$ Gpc simulation, where the total volume is $L^{3}$. 
Length scales are quoted under the assumption $h=0.704$ \citep[see][]{macpherson2018a}, and we use periodic boundary conditions in all simulations. The right panel of Figure~\ref{fig:thetarho} shows the density distribution at $z=0$, showing a two-dimensional slice through the midplane of the domain, normalised to the global average density, $\langle\rho\rangle_\mathrm{all}$. We evolve the matter distribution on a grid, treating dark matter as a fluid. This implies we cannot form virialised structures, and any dense regions will continue to collapse towards infinite density. This is a current limitation of any fully general relativistic cosmological simulation, since numerical relativity N-body codes for cosmology currently do not exist.

\subsection{Averaging}
It is common to compare the evolution of global averages in an inhomogeneous, anisotropic universe \citep{buchertehlers1997,buchert2000} to the evolution of a homogeneous, isotropic universe. However, the correct choice of averaging time-slice remains ambiguous due to the presence of nonlinearities. We adopt the averaging scheme of \citet{buchert2000}, generalised to any hypersurface of averaging \citep{larena2009b,brown2009a,brown2009b,clarkson2009,gasperini2010,umeh2011}. The average of a scalar function $\psi$ over a domain $\mathcal{D}$, located within the chosen hypersurface, is
\begin{equation}
	\langle\psi\rangle = \frac{1}{V_{\mathcal{D}}} \int_{\mathcal{D}}\psi \sqrt{\gamma}\;d^{3}X,
\end{equation}
where $V_{\mathcal{D}} = \int_{\mathcal{D}} \sqrt{\gamma} \,d^{3}X$ is the volume of the domain, with $\gamma$ the determinant of the spatial metric $\gamma_{ij}$.  We define our averaging hypersurfaces by observers with four-velocity $n_{\mu}=(-\alpha,0,0,0)$, where $\alpha$ is the lapse function, and we set the shift vector $\beta^{i}=0$. The four-velocity of these observers differs from the four-velocity of the fluid $u^{\mu}\equiv dx^{\mu}/d\tau$, where $\tau$ is the proper time.

\subsection{Measuring the Hubble parameter} \label{sec:hubble}
The local expansion rate of the fluid projected onto our averaging hypersurface is
\begin{equation} \label{eq:theta}
	\theta\equiv h^{\mu\nu}\nabla_{\mu}u_{\nu},
\end{equation}
where $h_{\mu\nu}\equiv g_{\mu\nu} + n_{\mu}n_{\nu}$, and $\nabla_{\mu}$ is the covariant derivative associated with the metric tensor $g_{\mu\nu}$. 
We define the effective Hubble parameter in a domain $\mathcal{D}$ to be
\begin{equation} \label{eq:hubbledef}
	\mathcal{H_D} \equiv \frac{1}{3}\langle\theta\rangle.
\end{equation}
In a Friedmann-Lema\^{i}tre-Robertson-Walker spacetime, \eqref{eq:hubbledef} reduces to the usual conformal Hubble parameter $\mathcal{H}=a'/a$, where $'$ represents a derivative with respect to conformal time. 

The local expansion rate is not necessarily what the observer measures. Observations of SNe \citep{riess2018a,riess2018b} measure the distance-redshift relation, and it is unclear how this relates to the local expansion rate. Recreating what an observer measures in an inhomogeneous Universe ultimately requires ray tracing \citep[see][]{giblin2016b,east2018}, which we leave to future work. 

\subsection{Averaging in subdomains}
In order to quantify $\mathcal{H_D}$ on different physical scales, we calculate averages over spherical subdomains placed randomly within the volume shown in Figure~\ref{fig:thetarho}. This allows us to analyse the effect of inhomogeneities independent of boundary effects. We calculate $\theta$ for each grid cell, and calculate $\mathcal{H_D}$ by averaging over subdomains of various radii $r_\mathcal{D}$.

Observations of SNe in the local universe span a redshift range of $0.023\lesssim z\lesssim0.15$ \citep{riess2011,riess2016,riess2018a,riess2018b}, corresponding to distances of $75\lesssim r_\mathcal{D} \lesssim 450\,h^{-1}$ Mpc \citep{wuhuterer2017,odderskov2014}. Local SNe with $z\lesssim0.023$ are excluded from the analysis in attempt to minimise cosmic variance; their inclusion results in a 3\% higher $H_0$, suggesting we are located in a void \citep{jha2007}. 

\begin{figure}
	\includegraphics[width=\columnwidth]{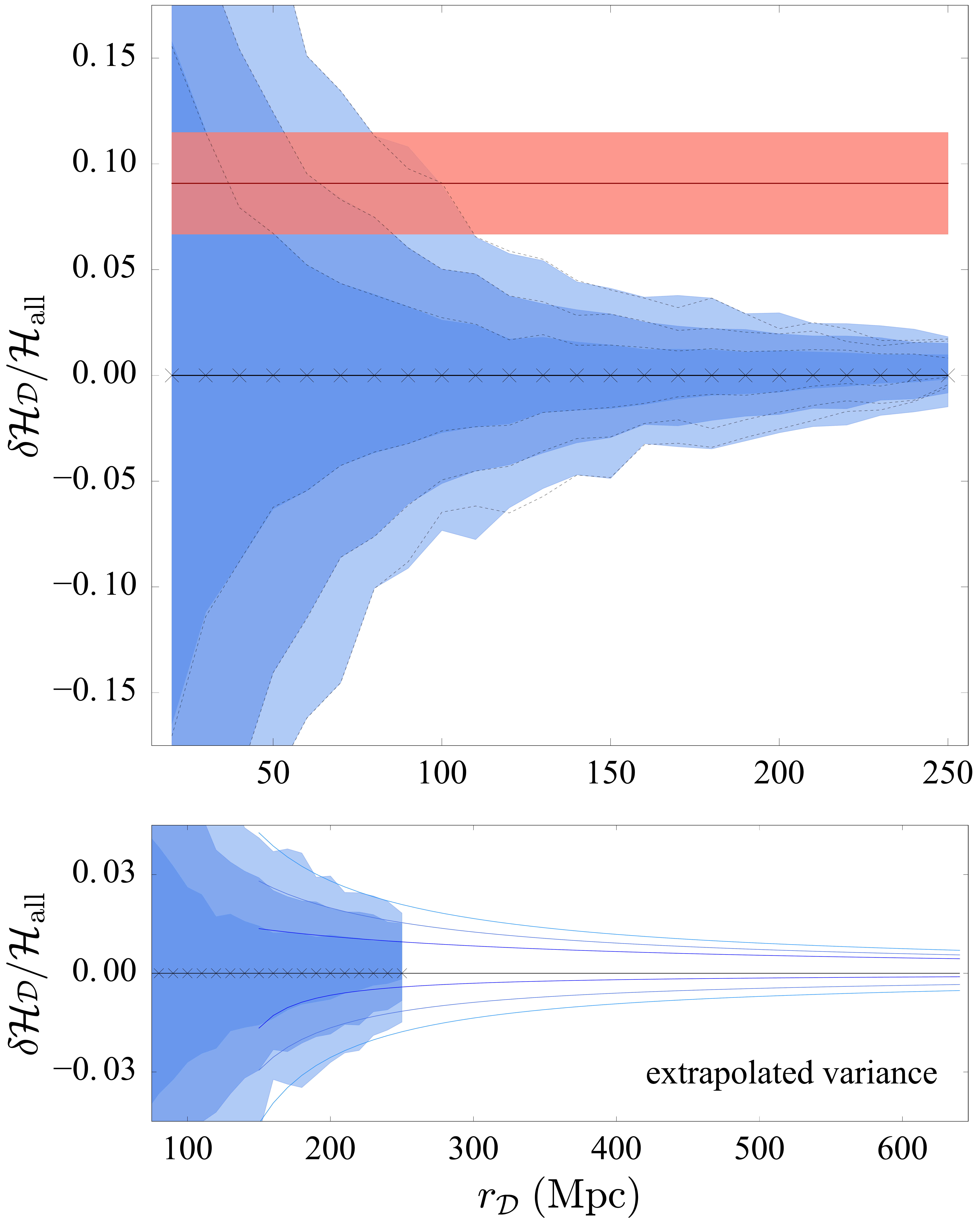}
    \caption{A general-relativistic measurement of $\mathcal{H_D}$ at $z=0$. Top panel is the fractional deviation measured in any one sphere from the average over the whole domain, $\mathcal{H}_\mathrm{all}$, as a function of averaging radius $r_{\mathcal{D}}$. Progressively lighter blue shaded regions are the 68\%, 95\% and 99.7\% confidence intervals, respectively. The red line is the measurement from \citet{riess2018b}, and the shaded region represents the 1$\sigma$ uncertainty. Dashed curves represent 68\%, 95\%, and 99.7\% confidence intervals for the same sample of spheres weighted as a function of redshift in accordance with the SNe sample used in \citet{riess2018a,riess2018b} \citep{wuhuterer2017,camarena2018}. Bottom panel shows the variance extrapolated to the full sample range \citep{riess2018a,riess2018b}. Progressively lighter blue curves are the extension of the 68\%, 95\%, and 99.7\% confidence intervals, respectively.}
    \label{fig:hubble}
\end{figure}

We approximate a measurement of the Hubble expansion using SNe by calculating the average local expansion rate over a variety of scales. We sample spherical regions with radii up to $r_\mathcal{D}=250$ Mpc to ensure individual spheres are sufficiently independent within our $L=1$ Gpc domain. We therefore calculate $\mathcal{H_D}$ on scales $75<r_\mathcal{D}<180\,h^{-1}$Mpc, corresponding to an effective survey range of $0.023\lesssim z\lesssim0.06$. The reduced range is due to the computational overhead of numerical relativity currently limiting us to domain sizes and resolutions of this order. We extrapolate to $r_\mathcal{D}=450\,h^{-1}$ Mpc to estimate the variance over the full range adopted in \citet{riess2018a,riess2018b}. We perform this extrapolation by fitting a function of the form $\delta\mathcal{H_D}/\mathcal{H}_\mathrm{all}\propto 1/r_\mathcal{D}$ using our calculated variance at $r_\mathcal{D}\geq150$ Mpc, to minimise the effect of small-scale fluctuations (see lower panel of Figure~\ref{fig:hubble}). To properly test the full range of observations, a larger simulation volume and resolution would be required. 


\section{Results} \label{sec:results}
The left panel of Figure~\ref{fig:thetarho} shows deviations in the Hubble parameter, relative to the global mean $\mathcal{H}_\mathrm{all}$, at $z=0$. We show a two-dimensional slice through the midplane of the $L=1$ Gpc domain. Green regions are expanding ($\theta>0$), while yellow to red regions are collapsing ($\theta<0$). This expansion is strongly correlated with the density field shown in the right panel, which displays filaments, voids, knots, and clusters. Due to our fluid treatment of dark matter, collapsing regions will continue to do so towards infinite density, implying all regions in the left panel of Figure~\ref{fig:thetarho} will average to the corresponding homogeneous expansion.

The top panel of Figure~\ref{fig:hubble} shows the deviation in the Hubble parameter as a function of averaging radius $r_\mathcal{D}$. Crosses represent the radii at which our calculations were done, and progressively lighter blue shaded regions represent the 65\%, 98\%, and 99.7\% confidence intervals over 1000 randomly placed spheres with the corresponding radius $r_\mathcal{D}$. The red line and shaded region show the mean and $1\sigma$ deviation of the \citet{riess2018b} measurement from the \citet{planck2018a} measurement, respectively. The bottom panel of Figure~\ref{fig:hubble} shows the 68\%, 95\%,and 99.7\% confidence contours (dark to light blue curves, respectively) extrapolated to the full redshift range used in \citet{riess2018a,riess2018b}. 

Considering our averaging spheres as a survey volume including SNe at redshifts $0.023\lesssim z\lesssim 0.06$, and assuming an isotropic distribution of objects across the sky with equal numbers of SNe at all redshift, we estimate the expected variance in a local $H_0$ measurement due to inhomogeneities as the variance in $\mathcal{H_D}$. We calculate the $\pm1\sigma$ variance in a measurement as the $84^{\mathrm{th}}$ and $16^{\mathrm{th}}$ percentiles of the full distribution of spheres sampled over the effective survey range, and similarly for the $2-3\sigma$ variance. Sampling all scales in the top panel of Figure~\ref{fig:hubble}, including local SNe with $z\lesssim0.023$, results in a $1\sigma$ variance of $\pm\,2.1\%$. Excluding these local SNe the variance drops to (+1.2,-1.1)\%. 
We extrapolate to the full survey range $0.023\lesssim z\lesssim 0.15$ (bottom panel of Figure~\ref{fig:hubble}) by fitting a function $\delta\mathcal{H_D}/\mathcal{H}_\mathrm{all}\propto 1/r_\mathcal{D}$ to each confidence contour in Figure~\ref{fig:hubble}. While not intended to be a precise measure of the variance at large scales, we estimate a $1\sigma$ variance of (+0.8,-0.4)\%.

\begin{figure*}
	\includegraphics[width=\textwidth]{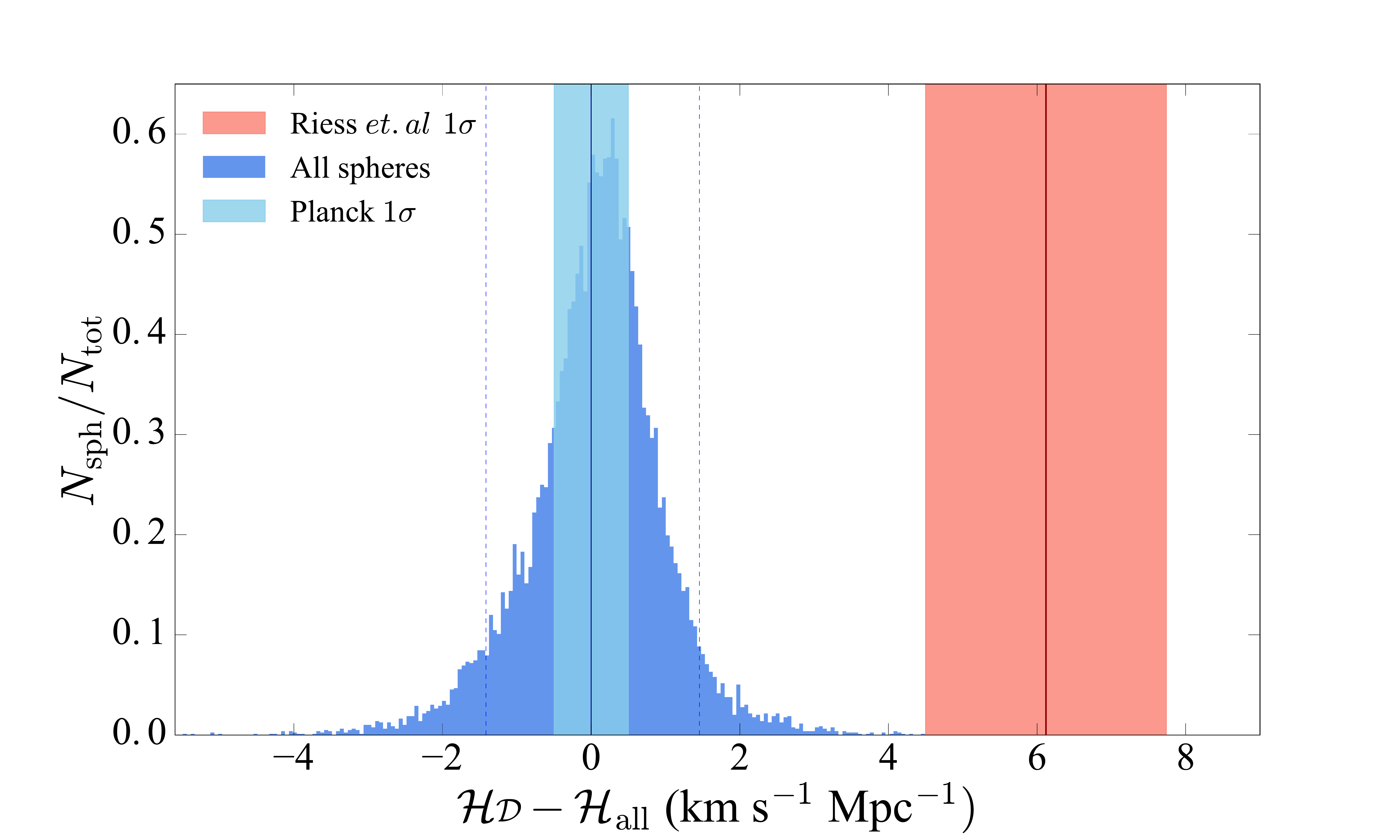}
    \caption{Local deviations in the Hubble parameter due to inhomogeneities. We show the full distribution of all spheres in the range $75< r_\mathcal{D} <180\,h^{-1}$ Mpc in blue. The dashed blue lines represent the $1\sigma$ deviation of the inhomogeneous distribution. The blue shaded region represents the $1\sigma$ uncertainties on the \citet{planck2016params} measurement, while the solid red line and shaded region represent the mean and $1\sigma$ deviation in the \citet{riess2018b} measurement, respectively.}
    \label{fig:hubble_dist}
\end{figure*}


The blue distribution in Figure~\ref{fig:hubble_dist} shows the local deviation in the Hubble parameter relative to the global mean, versus the fraction of total spheres with that deviation, $N_\mathrm{sph}/N_\mathrm{tot}$. We show the full sample of spheres in the range $0.023\lesssim z\lesssim0.06$, with the corresponding $1\sigma$ variations shown as dashed lines. The blue line and shaded region represent the \citet{planck2018a} measurement and $1\sigma$ uncertainties, respectively, while the red line and shaded region shows the \citet{riess2018b} measurement and the $1\sigma$ uncertainties, respectively. 

The Supercal SNe compilation \citep{scolnic2015}, used by \citet{riess2016}, does not contain equal numbers of SNe at all redshifts; a larger number of objects are sampled at low redshifts. Weighting our results in line with the redshift distribution of the sample \citep[as shown in][]{wuhuterer2017,camarena2018} we find the variance in the Hubble parameter increases to $(+1.5,-1.6)\%$ over our reduced redshift range. Dashed curves in the top panel of Figure~\ref{fig:hubble} show the variance as a function of averaging radius for the weighted sample. We proceed using the weighted sample for further analysis.

Extending to the $3\sigma$ variance over $0.023\lesssim z\lesssim 0.06$ we find a local Hubble constant can be up to 6.2\% larger than the mean. Taking the \citet{planck2018a} measurement of $67.4\pm 0.5$ km s$^{-1}$Mpc$^{-1}$ as the global mean expansion rate, this implies that if an observers position in the cosmic web is relatively underdense, she may measure a Hubble parameter up to $4.2$ km s$^{-1}$Mpc$^{-1}$ larger. Hence a local measurement using SNe could reach $H_0=71.6 \pm 1.62$ km s$^{-1}$Mpc$^{-1}$, assuming the same statistical uncertainties as \citet{riess2018b}. This measurement would then be in $2.5\sigma$ tension with \citet{planck2018a}. 

In order to completely resolve the tension between a local measurement and the global value, we must restrict our sample range to $60<r_\mathcal{D}<180\,h^{-1}$ Mpc, or $0.02\lesssim z\lesssim0.06$. Over these scales, our $3\sigma$ variance in the Hubble parameter implies a local $H_0$ measurement could be up to 8.7\%, or $5.9$ km s$^{-1}$Mpc$^{-1}$, larger than the global expansion. Again taking the \citet{planck2018a} value as the global expansion, a local measurement could reach $H_0=73.3 \pm 1.62$ km s$^{-1}$Mpc$^{-1}$ purely based on the observers location in an inhomogeneous universe. This is consistent with the \citet{riess2018b} measurement within $1\sigma$.

\section{discussion}\label{sec:discuss}

The variance in the effective Hubble parameter shown in Figure~\ref{fig:hubble} cannot resolve the tension between the \citet{planck2018a} and \citet{riess2018b} measurements. Excluding local SNe with $z\lesssim0.023$ we find the variance in the Hubble parameter due to inhomogeneities is (+1.5,-1.6)\% over a reduced redshift range. We find an observer can only measure a local Hubble parameter up to 8.7\% higher than the global value when \textit{further} reducing the survey range to $0.02\lesssim z\lesssim0.06$. The restricted range required for such a measurement emphasises that it is unlikely to completely resolve the tension by local variance in expansion rate. Extrapolating our results to the full survey range results in an expected variance below percent-level, however, as the precision of cosmological surveys continues to improve, variations of this size can be significant.

In \citet{macpherson2018a}, we analysed the effects of inhomogeneities on globally averaged quantities. We found global averages coincide with the equivalent homogeneous, isotropic model, with negligible backreaction effects on the global expansion. These results are subject to several caveats, which we outline below.

In our simulations we treat dark matter as a fluid, implying we cannot form virialised structures. Any structures that should have formed dark matter haloes will continue to collapse to a single point, eventually growing towards infinite density. Ideally, a particle method would be used for simulating dark matter as dust. We cannot directly compare our simulations to Newtonian N-body simulations due to this difference, in addition to gauge differences, however we can check for consistency of results. On scales $r_\mathcal{D}=50, 75$ and $100\,h^{-1}$Mpc we find variations of $\pm\,4.3\%$, $\pm\,2.4\%,$ and $(+1.1,-0.6)\%$, respectively. These are consistent with Newtonian predictions, also sampling observers randomly located in space, from \citet{wojtak2014} and \citet{odderskov2016} to within $\lesssim1\%$. However, to address whether this difference is due to general relativistic effects or computational differences, we ultimately require a particle treatment of dark matter alongside numerical relativity.


Our results may be considered an upper limit for the variance in the Hubble parameter over the scales we sample for several reasons. We assume averages over a purely spatial volume, when in reality an observer would measure their past light cone. As we look back in time, structures are more smoothed out, which would reduce the overall variance. In addition, we evolve our simulations assuming $\Lambda=0$; a matter-dominated universe at the initial instance. We do not fix $\Omega_m=1$ over the course of the simulation, however, globally we find $\Omega_m=1$ to within computational error for all time \citep{macpherson2018a}. This implies the growth rate, $f$, of structures in our simulation will be larger than in $\Lambda$CDM, since $f=\Omega_m^{0.55}$ \citep{linder2005}, resulting in a larger density contrast in general. This will also increase our variance in the Hubble parameter relative to that measured in the Universe where $\Omega_m\approx0.3$ is well constrained \citep[e.g.][]{DESCollab2017a,bonvin2017,planck2018a,bennett2013}.

The effects of inhomogeneities can be dependent on the choice of observers. \citet{adamek2017} used weak-field relativistic N-body simulations to study variance in the Hubble parameter in the comoving synchronous gauge and the Poisson gauge. In the comoving gauge the variance in the Hubble parameter reached 10\% at $z=0$, while the Poisson gauge remained below 0.01\%. A direct comparison to this work is not possible due to different definitions of the local expansion, however it outlines the importance of carefully choosing the averaging hypersurface. The comoving gauge is often used to represent observers on Earth, however this gauge breaks down at low redshifts due to shell crossings, and so it has been suggested the Poisson gauge -- similar to the gauge used here -- is better suited to study the effects of inhomogeneities in the nonlinear regime with simulations \citep{adamek2017}.


\section{conclusions}
We have investigated the effects of inhomogeneities on local measurements of the Hubble parameter. Using numerical relativity we have simulated the growth of density fluctuations drawn from the CMB through to $z=0$. We have calculated the expansion rate of dark matter within randomly placed spheres of various radii from a $256^{3}$ resolution simulation with domain size $L=1$ Gpc. Our conclusions are:
\begin{enumerate}
	\item We measure a (+1.5,-1.6)\% variance in the local expansion rate due to inhomogeneities over $0.023\lesssim z\lesssim0.06$ with a weighted sample of averaging spheres. 
	\item Estimating an extension to our results over $0.023\lesssim z\lesssim0.15$ reduces the variance to (+0.8,-0.4)\%. This is consistent with predictions from Newtonian N-body simulations. 
	\item Our $3\sigma$ variance in the Hubble parameter of 6.2\%, over $0.023\lesssim z\lesssim0.06$, could reduce the tension between a local and global measurement to $2.5\sigma$.
	\item When restricting the survey range to include more nearby SNe, the tension is resolved. Over scales $0.02\lesssim z\lesssim0.06$, a local calculation of $\mathcal{H_D}$ can be up to 8.7\% larger than the global value. However, since the \citet{riess2018a,riess2018b} measurement considers a significantly wider survey range, we conclude that the tension cannot be explained by local inhomogeneities under our assumptions.
\end{enumerate}

\acknowledgments
We thank the anonymous referee whose comments significantly improved the presentation of this work. We thank Chris Blake, Adam Riess, Krzysztof Bolejko, Marco Bruni, Syksy R\"{a}s\"{a}nen, Julian Adamek, Eloisa Bentivegna, Caitlin Adams, David Wiltshire, Chris Clarkson, Ruth Durrer, Timothy Clifton, and Tom Giblin for useful feedback and discussions, in general, as well as specific to this work. HJM especially thanks Marco Bruni and the University of Portsmouth for financial support and hospitality during the production of this work. HJM also thanks the organisers and participants of the Inhomogeneous Cosmologies conference in Toru\'n 2017 for feedback and support. This work was supported by resources provided by the Pawsey Supercomputing Centre with funding from the Australian Government and the Government of Western Australia. PDL is supported through Australian Research Council (ARC) Future Fellowship FT160100112 and ARC Discovery Project DP180103155.  DJP is supported through ARC FT130100034. 

\bibliographystyle{aasjournal}
\bibliography{litreview}

\begin{thebibliography}{}
\expandafter\ifx\csname natexlab\endcsname\relax\def\natexlab#1{#1}\fi
\providecommand{\url}[1]{\href{#1}{#1}}
\providecommand{\dodoi}[1]{doi:~\href{http://doi.org/#1}{\nolinkurl{#1}}}
\providecommand{\doeprint}[1]{\href{http://ascl.net/#1}{\nolinkurl{http://ascl.net/#1}}}
\providecommand{\doarXiv}[1]{\href{https://arxiv.org/abs/#1}{\nolinkurl{https://arxiv.org/abs/#1}}}

\bibitem[{{Adamek} {et~al.}(2017){Adamek}, {Clarkson}, {Daverio}, {Durrer}, \&
  {Kunz}}]{adamek2017}
{Adamek}, J., {Clarkson}, C., {Daverio}, D., {Durrer}, R., \& {Kunz}, M. 2017,
  ArXiv e-prints.
\newblock \doarXiv{1706.09309}

\bibitem[{{Addison} {et~al.}(2016){Addison}, {Huang}, {Watts}, {Bennett},
  {Halpern}, {Hinshaw}, \& {Weiland}}]{addison2016}
{Addison}, G.~E., {Huang}, Y., {Watts}, D.~J., {et~al.} 2016, \apj, 818, 132,
  \dodoi{10.3847/0004-637X/818/2/132}

\bibitem[{{Ben-Dayan} {et~al.}(2014){Ben-Dayan}, {Durrer}, {Marozzi}, \&
  {Schwarz}}]{ben-dayan2014}
{Ben-Dayan}, I., {Durrer}, R., {Marozzi}, G., \& {Schwarz}, D.~J. 2014,
  Physical Review Letters, 112, 221301, \dodoi{10.1103/PhysRevLett.112.221301}

\bibitem[{{Bennett} {et~al.}(2013){Bennett}, {Larson}, {Weiland}, {Jarosik},
  {Hinshaw}, {Odegard}, {Smith}, {Hill}, {Gold}, {Halpern}, {Komatsu}, {Nolta},
  {Page}, {Spergel}, {Wollack}, {Dunkley}, {Kogut}, {Limon}, {Meyer}, {Tucker},
  \& {Wright}}]{bennett2013}
{Bennett}, C.~L., {Larson}, D., {Weiland}, J.~L., {et~al.} 2013, \apj, 208, 20,
  \dodoi{10.1088/0067-0049/208/2/20}

\bibitem[{{Bentivegna} \& {Bruni}(2016)}]{bentivegna2016a}
{Bentivegna}, E., \& {Bruni}, M. 2016, Physical Review Letters, 116, 251302,
  \dodoi{10.1103/PhysRevLett.116.251302}

\bibitem[{{Bolejko}(2017)}]{bolejko2017b}
{Bolejko}, K. 2017, ArXiv e-prints.
\newblock \doarXiv{1707.01800}

\bibitem[{{Bolejko}(2018)}]{bolejko2018b}
---. 2018, \prd, 97, 103529, \dodoi{10.1103/PhysRevD.97.103529}

\bibitem[{{Bonvin} {et~al.}(2017){Bonvin}, {Courbin}, {Suyu}, {Marshall},
  {Rusu}, {Sluse}, {Tewes}, {Wong}, {Collett}, {Fassnacht}, {Treu}, {Auger},
  {Hilbert}, {Koopmans}, {Meylan}, {Rumbaugh}, {Sonnenfeld}, \&
  {Spiniello}}]{bonvin2017}
{Bonvin}, V., {Courbin}, F., {Suyu}, S.~H., {et~al.} 2017, \mnras, 465, 4914,
  \dodoi{10.1093/mnras/stw3006}

\bibitem[{{Brown} {et~al.}(2009{\natexlab{a}}){Brown}, {Behrend}, \&
  {Malik}}]{brown2009b}
{Brown}, I.~A., {Behrend}, J., \& {Malik}, K.~A. 2009{\natexlab{a}}, \jcap, 11,
  027, \dodoi{10.1088/1475-7516/2009/11/027}

\bibitem[{{Brown} {et~al.}(2009{\natexlab{b}}){Brown}, {Robbers}, \&
  {Behrend}}]{brown2009a}
{Brown}, I.~A., {Robbers}, G., \& {Behrend}, J. 2009{\natexlab{b}}, \jcap, 4,
  016, \dodoi{10.1088/1475-7516/2009/04/016}

\bibitem[{{Buchert} \& {Ehlers}(1997)}]{buchertehlers1997}
{Buchert}, T., \& {Ehlers}, J. 1997, \aap, 320, 1

\bibitem[{{Buchert} {et~al.}(2000){Buchert}, {Kerscher}, \&
  {Sicka}}]{buchert2000}
{Buchert}, T., {Kerscher}, M., \& {Sicka}, C. 2000, \prd, 62, 043525,
  \dodoi{10.1103/PhysRevD.62.043525}

\bibitem[{{Buchert} {et~al.}(2015){Buchert}, {Carfora}, {Ellis}, {Kolb},
  {MacCallum}, {Ostrowski}, {R{\"a}s{\"a}nen}, {Roukema}, {Andersson}, {Coley},
  \& {Wiltshire}}]{buchert2015}
{Buchert}, T., {Carfora}, M., {Ellis}, G.~F.~R., {et~al.} 2015, Classical and
  Quantum Gravity, 32, 215021, \dodoi{10.1088/0264-9381/32/21/215021}

\bibitem[{{Camarena} \& {Marra}(2018)}]{camarena2018}
{Camarena}, D., \& {Marra}, V. 2018, ArXiv e-prints.
\newblock \doarXiv{1805.09900}

\bibitem[{{Clarkson} {et~al.}(2009){Clarkson}, {Ananda}, \&
  {Larena}}]{clarkson2009}
{Clarkson}, C., {Ananda}, K., \& {Larena}, J. 2009, \prd, 80, 083525,
  \dodoi{10.1103/PhysRevD.80.083525}

\bibitem[{{Cusin} {et~al.}(2017){Cusin}, {Pitrou}, \& {Uzan}}]{cusin2017}
{Cusin}, G., {Pitrou}, C., \& {Uzan}, J.-P. 2017, \jcap, 3, 038,
  \dodoi{10.1088/1475-7516/2017/03/038}

\bibitem[{{DES Collaboration} {et~al.}(2017){DES Collaboration}, {Abbott},
  {Abdalla}, {Alarcon}, {Aleksi{\'c}}, {Allam}, {Allen}, {Amara}, {Annis},
  {Asorey}, {Avila}, {Bacon}, {et~al.}}]{DESCollab2017a}
{DES Collaboration}, {Abbott}, T.~M.~C., {Abdalla}, F.~B., {et~al.} 2017, ArXiv
  e-prints.
\newblock \doarXiv{1708.01530}

\bibitem[{{Dhawan} {et~al.}(2018){Dhawan}, {Jha}, \&
  {Leibundgut}}]{dhawan2018a}
{Dhawan}, S., {Jha}, S.~W., \& {Leibundgut}, B. 2018, \aap, 609, A72,
  \dodoi{10.1051/0004-6361/201731501}

\bibitem[{{East} {et~al.}(2018){East}, {Wojtak}, \& {Abel}}]{east2018}
{East}, W.~E., {Wojtak}, R., \& {Abel}, T. 2018, \prd, 97, 043509,
  \dodoi{10.1103/PhysRevD.97.043509}

\bibitem[{{Efstathiou}(2014)}]{efstathiou2014}
{Efstathiou}, G. 2014, \mnras, 440, 1138, \dodoi{10.1093/mnras/stu278}

\bibitem[{{Gasperini} {et~al.}(2010){Gasperini}, {Marozzi}, \&
  {Veneziano}}]{gasperini2010}
{Gasperini}, M., {Marozzi}, G., \& {Veneziano}, G. 2010, \jcap, 2, 009,
  \dodoi{10.1088/1475-7516/2010/02/009}

\bibitem[{{Giblin} {et~al.}(2016{\natexlab{a}}){Giblin}, {Mertens}, \&
  {Starkman}}]{giblin2016a}
{Giblin}, J.~T., {Mertens}, J.~B., \& {Starkman}, G.~D. 2016{\natexlab{a}},
  Physical Review Letters, 116, 251301, \dodoi{10.1103/PhysRevLett.116.251301}

\bibitem[{{Giblin} {et~al.}(2016{\natexlab{b}}){Giblin}, {Mertens}, \&
  {Starkman}}]{giblin2016b}
{Giblin}, Jr., J.~T., {Mertens}, J.~B., \& {Starkman}, G.~D.
  2016{\natexlab{b}}, \apj, 833, 247, \dodoi{10.3847/1538-4357/833/2/247}

\bibitem[{{Giblin} {et~al.}(2017){Giblin}, {Mertens}, \&
  {Starkman}}]{giblin2017a}
---. 2017, Classical and Quantum Gravity, 34, 214001,
  \dodoi{10.1088/1361-6382/aa8af9}

\bibitem[{{Hoscheit} \& {Barger}(2018)}]{hoscheit2018}
{Hoscheit}, B.~L., \& {Barger}, A.~J. 2018, \apj, 854, 46,
  \dodoi{10.3847/1538-4357/aaa59b}

\bibitem[{{Jha} {et~al.}(2007){Jha}, {Riess}, \& {Kirshner}}]{jha2007}
{Jha}, S., {Riess}, A.~G., \& {Kirshner}, R.~P. 2007, \apj, 659, 122,
  \dodoi{10.1086/512054}

\bibitem[{{Larena}(2009)}]{larena2009b}
{Larena}, J. 2009, \prd, 79, 084006, \dodoi{10.1103/PhysRevD.79.084006}

\bibitem[{{Lewis} \& {Bridle}(2002)}]{lewis2002}
{Lewis}, A., \& {Bridle}, S. 2002, \prd, 66, 103511,
  \dodoi{10.1103/PhysRevD.66.103511}

\bibitem[{{Linder}(2005)}]{linder2005}
{Linder}, E.~V. 2005, \prd, 72, 043529, \dodoi{10.1103/PhysRevD.72.043529}

\bibitem[{{Macpherson} {et~al.}(2018){Macpherson}, {Price}, \&
  {Lasky}}]{macpherson2018a}
{Macpherson}, H., {Price}, D.~J., \& {Lasky}, P.~D. 2018, ArXiv e-prints.
\newblock \doarXiv{1807.01711}

\bibitem[{{Macpherson} {et~al.}(2017){Macpherson}, {Lasky}, \&
  {Price}}]{macpherson2017a}
{Macpherson}, H.~J., {Lasky}, P.~D., \& {Price}, D.~J. 2017, \prd, 95, 064028,
  \dodoi{10.1103/PhysRevD.95.064028}

\bibitem[{{Marra} {et~al.}(2013){Marra}, {Amendola}, {Sawicki}, \&
  {Valkenburg}}]{marra2013}
{Marra}, V., {Amendola}, L., {Sawicki}, I., \& {Valkenburg}, W. 2013, Physical
  Review Letters, 110, 241305, \dodoi{10.1103/PhysRevLett.110.241305}

\bibitem[{{Odderskov} {et~al.}(2014){Odderskov}, {Hannestad}, \&
  {Haugb{\o}lle}}]{odderskov2014}
{Odderskov}, I., {Hannestad}, S., \& {Haugb{\o}lle}, T. 2014, \jcap, 10, 028,
  \dodoi{10.1088/1475-7516/2014/10/028}

\bibitem[{{Odderskov} {et~al.}(2016){Odderskov}, {Koksbang}, \&
  {Hannestad}}]{odderskov2016}
{Odderskov}, I., {Koksbang}, S.~M., \& {Hannestad}, S. 2016, \jcap, 2, 001,
  \dodoi{10.1088/1475-7516/2016/02/001}

\bibitem[{{Ostrowski} {et~al.}(2013){Ostrowski}, {Roukema}, \&
  {Buchert}}]{ostrowski2013}
{Ostrowski}, J.~J., {Roukema}, B.~F., \& {Buchert}, T. 2013, ArXiv e-prints.
\newblock \doarXiv{1311.5402}

\bibitem[{{Planck Collaboration} {et~al.}(2016){Planck Collaboration}, {Ade},
  {Aghanim}, {Arnaud}, {Ashdown}, {Aumont}, {Baccigalupi}, {Banday},
  {Barreiro}, {Bartlett}, \& et~al.}]{planck2016params}
{Planck Collaboration}, {Ade}, P.~A.~R., {Aghanim}, N., {et~al.} 2016, \aap,
  594, A13, \dodoi{10.1051/0004-6361/201525830}

\bibitem[{{Planck Collaboration} {et~al.}(2018){Planck Collaboration},
  {Aghanim}, {Akrami}, {Ashdown}, {Aumont}, {Baccigalupi}, {Ballardini},
  {Banday}, {Barreiro}, {Bartolo}, {Basak}, {Battye}, {Benabed}, {Bernard},
  {Bersanelli}, {Bielewicz}, {Bock}, {Bond}, {Borrill}, {Bouchet}, {Boulanger},
  {Bucher}, {Burigana}, {Butler}, {Calabrese}, {Cardoso}, {Carron},
  {Challinor}, {Chiang}, {Chluba}, {Colombo}, {Combet}, {Contreras}, {Crill},
  {Cuttaia}, {de Bernardis}, {de Zotti}, {Delabrouille}, {Delouis}, {Di
  Valentino}, {Diego}, {Dor{\'e}}, {Douspis}, {Ducout}, {Dupac}, {Dusini},
  {Efstathiou}, {Elsner}, {En{\ss}lin}, {Eriksen}, {Fantaye}, {Farhang},
  {Fergusson}, {Fernandez-Cobos}, {Finelli}, {Forastieri}, {Frailis},
  {Franceschi}, {Frolov}, {Galeotta}, {Galli}, {Ganga}, {G{\'e}nova-Santos},
  {Gerbino}, {Ghosh}, {Gonz{\'a}lez-Nuevo}, {G{\'o}rski}, {Gratton},
  {Gruppuso}, {Gudmundsson}, {Hamann}, {Handley}, {Herranz}, {Hivon}, {Huang},
  {Jaffe}, {Jones}, {Karakci}, {Keih{\"a}nen}, {Keskitalo}, {Kiiveri}, {Kim},
  {Kisner}, {Knox}, {Krachmalnicoff}, {Kunz}, {Kurki-Suonio}, {Lagache},
  {Lamarre}, {Lasenby}, {Lattanzi}, {Lawrence}, {Le Jeune}, {Lemos},
  {Lesgourgues}, {Levrier}, {Lewis}, {Liguori}, {Lilje}, {Lilley}, {Lindholm},
  {L{\'o}pez-Caniego}, {Lubin}, {Ma}, {Mac{\'{\i}}as-P{\'e}rez}, {Maggio},
  {Maino}, {Mandolesi}, {Mangilli}, {Marcos-Caballero}, {Maris}, {Martin},
  {Martinelli}, {Mart{\'{\i}}nez-Gonz{\'a}lez}, {Matarrese}, {Mauri}, {McEwen},
  {Meinhold}, {Melchiorri}, {Mennella}, {Migliaccio}, {Millea}, {Mitra},
  {Miville-Desch{\^e}nes}, {Molinari}, {Montier}, {Morgante}, {Moss}, {Natoli},
  {N{\o}rgaard-Nielsen}, {Pagano}, {Paoletti}, {Partridge}, {Patanchon},
  {Peiris}, {Perrotta}, {Pettorino}, {Piacentini}, {Polastri}, {Polenta},
  {Puget}, {Rachen}, {Reinecke}, {Remazeilles}, {Renzi}, {Rocha}, {Rosset},
  {Roudier}, {Rubi{\~n}o-Mart{\'{\i}}n}, {Ruiz-Granados}, {Salvati}, {Sandri},
  {Savelainen}, {Scott}, {Shellard}, {Sirignano}, {Sirri}, {Spencer},
  {Sunyaev}, {Suur-Uski}, {Tauber}, {Tavagnacco}, {Tenti}, {Toffolatti},
  {Tomasi}, {Trombetti}, {Valenziano}, {Valiviita}, {Van Tent}, {Vibert},
  {Vielva}, {Villa}, {Vittorio}, {Wandelt}, {Wehus}, {White}, {White},
  {Zacchei}, \& {Zonca}}]{planck2018a}
{Planck Collaboration}, {Aghanim}, N., {Akrami}, Y., {et~al.} 2018, ArXiv
  e-prints.
\newblock \doarXiv{1807.06209}

\bibitem[{{R{\"a}s{\"a}nen}(2006)}]{rasanen2006b}
{R{\"a}s{\"a}nen}, S. 2006, \jcap, 11, 3, \dodoi{10.1088/1475-7516/2006/11/003}

\bibitem[{{R{\"a}s{\"a}nen}(2008)}]{rasanen2008}
---. 2008, \jcap, 4, 026, \dodoi{10.1088/1475-7516/2008/04/026}

\bibitem[{{Riess} {et~al.}(2011){Riess}, {Macri}, {Casertano}, {Lampeitl},
  {Ferguson}, {Filippenko}, {Jha}, {Li}, \& {Chornock}}]{riess2011}
{Riess}, A.~G., {Macri}, L., {Casertano}, S., {et~al.} 2011, \apj, 730, 119,
  \dodoi{10.1088/0004-637X/730/2/119}

\bibitem[{{Riess} {et~al.}(2016){Riess}, {Macri}, {Hoffmann}, {Scolnic},
  {Casertano}, {Filippenko}, {Tucker}, {Reid}, {Jones}, {Silverman},
  {Chornock}, {Challis}, {Yuan}, {Brown}, \& {Foley}}]{riess2016}
{Riess}, A.~G., {Macri}, L.~M., {Hoffmann}, S.~L., {et~al.} 2016, \apj, 826,
  56, \dodoi{10.3847/0004-637X/826/1/56}

\bibitem[{{Riess} {et~al.}(2018{\natexlab{a}}){Riess}, {Casertano}, {Yuan},
  {Macri}, {Bucciarelli}, {Lattanzi}, {MacKenty}, {Bowers}, {Zheng},
  {Filippenko}, {Huang}, \& {Anderson}}]{riess2018b}
{Riess}, A.~G., {Casertano}, S., {Yuan}, W., {et~al.} 2018{\natexlab{a}}, ArXiv
  e-prints.
\newblock \doarXiv{1804.10655}

\bibitem[{{Riess} {et~al.}(2018{\natexlab{b}}){Riess}, {Casertano}, {Yuan},
  {Macri}, {Anderson}, {MacKenty}, {Bowers}, {Clubb}, {Filippenko}, {Jones}, \&
  {Tucker}}]{riess2018a}
---. 2018{\natexlab{b}}, \apj, 855, 136, \dodoi{10.3847/1538-4357/aaadb7}

\bibitem[{{Roy} {et~al.}(2011){Roy}, {Buchert}, {Carloni}, \&
  {Obadia}}]{roy2011}
{Roy}, X., {Buchert}, T., {Carloni}, S., \& {Obadia}, N. 2011, Classical and
  Quantum Gravity, 28, 165004, \dodoi{10.1088/0264-9381/28/16/165004}

\bibitem[{{Scolnic} {et~al.}(2015){Scolnic}, {Casertano}, {Riess}, {Rest},
  {Schlafly}, {Foley}, {Finkbeiner}, {Tang}, {Burgett}, {Chambers}, {Draper},
  {Flewelling}, {Hodapp}, {Huber}, {Kaiser}, {Kudritzki}, {Magnier},
  {Metcalfe}, \& {Stubbs}}]{scolnic2015}
{Scolnic}, D., {Casertano}, S., {Riess}, A., {et~al.} 2015, \apj, 815, 117,
  \dodoi{10.1088/0004-637X/815/2/117}

\bibitem[{{Shi} \& {Turner}(1998)}]{shi1998}
{Shi}, X., \& {Turner}, M.~S. 1998, \apj, 493, 519, \dodoi{10.1086/305169}

\bibitem[{{Sundell} {et~al.}(2015){Sundell}, {M{\"o}rtsell}, \&
  {Vilja}}]{sundell2015}
{Sundell}, P., {M{\"o}rtsell}, E., \& {Vilja}, I. 2015, \jcap, 8, 037,
  \dodoi{10.1088/1475-7516/2015/08/037}

\bibitem[{{Umeh} {et~al.}(2011){Umeh}, {Larena}, \& {Clarkson}}]{umeh2011}
{Umeh}, O., {Larena}, J., \& {Clarkson}, C. 2011, \jcap, 3, 029,
  \dodoi{10.1088/1475-7516/2011/03/029}

\bibitem[{{Wojtak} {et~al.}(2014){Wojtak}, {Knebe}, {Watson}, {Iliev},
  {He{\ss}}, {Rapetti}, {Yepes}, \& {Gottl{\"o}ber}}]{wojtak2014}
{Wojtak}, R., {Knebe}, A., {Watson}, W.~A., {et~al.} 2014, \mnras, 438, 1805,
  \dodoi{10.1093/mnras/stt2321}

\bibitem[{{Wu} \& {Huterer}(2017)}]{wuhuterer2017}
{Wu}, H.-Y., \& {Huterer}, D. 2017, \mnras, 471, 4946,
  \dodoi{10.1093/mnras/stx1967}

\bibitem[{{Zehavi} {et~al.}(1998){Zehavi}, {Riess}, {Kirshner}, \&
  {Dekel}}]{zehavi1998}
{Zehavi}, I., {Riess}, A.~G., {Kirshner}, R.~P., \& {Dekel}, A. 1998, \apj,
  503, 483, \dodoi{10.1086/306015}

\end{thebibliography}



\end{document}